\documentclass[aps,prd,11pt,superscriptaddress,nofootinbib]{revtex4-2}

\usepackage[utf8]{inputenc}
\usepackage{graphicx}
\usepackage{bm}
\usepackage{amsmath,amssymb}
\usepackage{latexsym}
\usepackage{xcolor}

\usepackage{amssymb}
\usepackage{tabularx}
\usepackage{nicematrix}

\begin{document}

\title{Accurate quasinormal modes of the analogue black holes}

\author{Jerzy Matyjasek}\email{jurek@kft.umcs.lublin.pl, jirinek@gmail.com}
\author{Kristian Benda}\email{kriben@gmail.com}
\author{Maja Stafińska}\email{majastafinska@gmail.com}
\affiliation{Institute of Physics,
Maria Curie-Sk\l odowska University\\
pl. Marii Curie-Sk\l odowskiej 1,
20-031 Lublin, Poland}

\begin{abstract}

We study the quasinormal modes of the spherically-symmetric $(2+1)$-dimensional 
analogue black
hole, modeled by the ``draining bathtub'' fluid flow, and the $(3+1)$-dimensional
canonical acoustic black hole. In the both cases the emphasis is on the accuracy.
Formally, the radial equation describing perturbations of the $(2+1)$-dimensional black hole is
a special case of the general master equation of the 5-dimensional Tangherlini black
hole. Similarly, the $(3+1)$-dimensional equation can be obtained from the master
equation of the 7-dimensional Tangherlini black hole.  For the $(2+1)$-dimensional
analogue black hole we used three major techniques: the higher-order WKB method with
the Pad\'e summation, the Hill-determinant method  and the continued fraction method,
the latter two with the convergence acceleration. In the $(3+1)$-dimensional case, we
propose the simpler recurrence relations and explicitly demonstrate  that both
recurrences, i.e., the eight-term and the six-term recurrences yield identical
results. Since the application of the continued-fraction method require five (or
three) consecutive Gauss eliminations, we decided not to use this technique in the
$(3+1)$-dimensional case.  Instead, we used the Hill-determinant method in the two
incarnations and the higher-order WKB. 
We accept the results of our calculations if at least two (algorithmically) independent 
methods give the same answer to some prescribed accuracy. 
Our results correct and extend the results
existing in the literature and we believe that we approached assumed accuracy of 9
decimal places. In most cases, there is perfect agreement between all the methods; 
however, in a few cases, the performance of the higher-order WKB method is slightly 
worse.

\end{abstract}
\maketitle

\section{Introduction}

With the first detection of the gravitational waves, our hopes for opening of a
new observational window increased significantly.  And although currently we can only
draw very general conclusions from the received signals, it is conceivable that the
planned new generations of more sensitive detectors will provide us with detailed
information regarding the nature of the astrophysical sources of the gravitational
waves and the gravitational waves themselves.  However, it is also conceivable that
extraction of the imprints of some concrete physical phenomena from the received
signals may prove challenging. Therefore, while waiting for a significant
breakthrough in the observational techniques and the next generations of the
gravitational wave detectors, we should also focus on the experiments that can be, 
in principle, carried out in the laboratory. As the most common sources of detectable 
gravitational waves are black holes, it is natural that we should try to devise 
experimental setups for configurations that mimic astrophysical black holes.

Black holes are important components of many of our modern theories.
Indeed, they play a crucial role in explaining the formation of galaxies,
the creation and dynamics of accretion disks, active galactic nuclei,
and gamma-ray bursts, to name just a few examples. They may also lead
to some observable optical phenomena and (probably practically undetectable) Hawking radiation.
Since Unruh's seminal paper on black hole evaporation in a laboratory~\cite{Unruh}, there
has been a flurry of research on the so called analogue black holes and the interested
reader is referred to the excellent review ~\cite{livingrr}, research papers 
(see for example Refs.~\cite{rich1,rich2}) and the numerous references cited therein.
In this article we shall concentrate on the two models of the analogue black holes:
(2+1)-dimensional acoustic black hole (also called the draining bathtub model) and
(3+1)-dimensional canonical acoustic black hole~\cite{Visser_1998}.  Our task will be
to determine complex frequencies of the quasinormal modes of such configurations.

Quasinormal modes are radiated in response to a weak disturbance or when the initial 
strong disturbance becomes weak. In all these cases, all information regarding the nature of the initial 
disturbances is washed out, and the black hole radiates its own characteristic 
frequencies~\cite{KokkotasR,BertiR,KonoplyaR,NollertR}.
Although both types of the analogue black holes have been considered 
earlier~\cite{becarle,jose_hydro_1}, 
our objective is to extend the analysis of the quasinormal oscillations to higher 
fundamental modes and their overtones and to deliver highly accurate results 
that refine or correct the existing results in the literature. To accomplish 
this, depending on the specific mode, we have employed several methods, 
such as higher-order WKB with Padé summation\footnote{The analytic form of the WKB terms
is known up to the 16th order (see Refs.~\cite{jaOP,jago}). By higher-order WKB,
we mean numerically calculated terms of at least the 50th-order. }, the Hill determinant method 
with convergence acceleration, and, additionally, in the case 
of the (2+1)-dimensional acoustic black hole, the continued fraction method
also with convergence acceleration. Moreover, in the case of the canonical 
acoustic black hole, the Hill determinant method was used with two different 
recurrence relations. Since we have restricted ourselves to the three aforementioned
methods, some promising approaches, such as the asymptotic iteration 
method~\cite{wade_naylor} or the spectral method~\cite{Jansen}, were not employed.

Besides the fundamental modes, we have also determined a certain number of overtones
(labeled by the overtone number $n \geq 1$), not only to
capture the trend of the quasinormal frequency changes as $n$ increases, but also
because their calculation serves as the unique stress test for the adapted methods and algorithms.  
We accept the result of our calculations if at least two independent methods give 
the same answer to 9 decimal places, and if the results calculated using other competing 
methods considered in this paper differ from them only slightly.
In doing so we have two goals in mind. First, the
signal from the laboratory acoustic black holes will be contaminated and to extract
the quasinormal frequency it is  desirable to have at one's disposal the highly
accurate theoretical predictions.  Secondly, by demanding accuracy, one can compare
the actual performance of various methods, analyze their strengths and weaknesses,
and identify possible bottlenecks. The latter observation may be crucial in the case 
of `difficult' potentials, where methods relying on the higher-order  WKB-Pad\'e
approximation (and its variants) are especially valuable due to their universality and simplicity.
   
The paper is organized as follows. In Sec.~\ref{sekcja2} we briefly review the most
important for our calculations features of  the  $(2+1)$ and $(3+1)$-dimensional
acoustic black holes and introduce the radial equations describing perturbations. In
Sec.~\ref{sekcja3} the relation between the master equation of the $D$-dimensional
Schwarzschild-Tangherlini and radial equations of the $(2+1)$-dimensional and
$(3+1)$-dimensional acoustic black holes is investigated. In the case of 
the $(3+1)$-dimensional acoustic black hole, we also propose a slightly more 
complex series solution to the radial equation, which, nevertheless, yields 
a simpler six-term recurrence. (The coefficient of the recurrence relations 
for the $D$-dimensional black
hole are relegated to the appendix).  In Secs.~\ref{sekcja4} and \ref{sekcja5} we
describe our ideas and strategies for calculating the complex frequencies of the
quasinormal modes of the acoustic black holes. In Sec.~\ref{sekcja6}, we present the
results of our high-accuracy calculations and compare them with the ones previously 
published. The final discussion and conclusions are given in Sec.~\ref{sekcja7}.
Throughout the paper, we shall follow the terminology used in the black hole physics.

\section{(2+1)-dimensional and (3+1)-dimensional acoustic black holes}
\label{sekcja2} 
\subsection{(2+1)-dimensional black holes}

The line element of the static (2+1)-dimensional acoustic black hole 
(also called the draining bathtub model)  can be written in the form~\cite{Visser_1998,becarle}
\begin{equation}
 ds^{2} = -\left(c_{s}^{2} -\frac{A^{2} }{r^{2}}\right) d t^{2} 
         + \frac{2 A}{r} d t  d r  + d r^{2} + r^{2} d \phi^{2},
 \label{threed}
\end{equation}
where $A$ is a positive constant related to the radial velocity of the background fluid
\begin{equation}
v = - \frac{A}{r}
\end{equation}
and $c_{s}$ is the speed of sound.  It describes $(2 + 1)$-dimensional flow with a sink at the origin.
The fluid
is locally irrotational, (i.e., vorticity free), barotropic and inviscid.  
The acoustic event horizon is located at $r_{H} = A/c_{s}.$ 

The propagation of sound waves in the acoustic geometry is
described by a massless minimally coupled scalar field equation 
\begin{equation}
g^{ab} \nabla_{a} \nabla_{n} F = 0, 
\end{equation} 
which, for the line element (\ref{threed}), can be separated using simple substitution 
\begin{equation}
F(t,r,\phi) = \psi(r) e^{i(m\phi - \omega t)} .  
\end{equation} 
Now, rescaling the radial coordinate and the frequency according to 
$r \to r c_{s}/A$ and 
$\omega \to \omega c_{s}^{2}/A$, 
after some  manipulations~\cite{Visser_1998,becarle}, one obtains
\begin{equation}
\frac{d^{2}}{d r_{\star}^{2}} \psi + \left( \omega^{2} - V(r) \right) \psi = 0,
\label{master} 
\end{equation} 
where 
\begin{equation} 
V(r) = \left( 1 -
\frac{1}{r^{2}} \right)\left[\frac{1}{r^{2}} \left(m^{2} - \frac{1}{4} \right) +
\frac{5}{4 r^{4}} \right], \label{hy3} 
\end{equation}
$r_{\star}$ is the standard tortoise radial coordinate, $m$ is the azimuthal number,
 and $r_{H} =1.$

\subsection{(3+1)-dimensional black holes}

The line element describing the static $(3+1)$-dimensional nonrotating  acoustic black
hole~\cite{Visser_1998}  has the following form (for the discussion and
explanations see e.g., Ref.~\cite{becarle})
\begin{equation}
 ds^{2} = - c_{s}^{2}\left(1-\frac{r_{0}^{4}}{r^{4}} \right) d t^{2} +  
  \left(1-\frac{r_{0}^{4}}{r^{4}} \right)^{-1} d r^{2}
   + r^{2} \left(d \theta^{2} + \sin^{2}\theta\,d\phi^{2}  \right),
 \label{dim4}
\end{equation}
where $c_{s}$ is the speed of sound and $r_{0}$ is the normalization constant.
It is the acoustic metric related to the spherically-symmetric flow of incompressible barotropic fluid.
The time-dependent version of this  metric may be constructed around spherically-symmetric
buble with oscillating radius (see~\cite{Visser_1998} and the references therein).

The propagation of small disturbances is described by the  minimally coupled
massless Klein-Gordon equation, which, because of the symmetries, can easily be
separated using
\begin{equation}
 \psi(t,r,\theta,\phi) = \frac{\psi(r)}{r} e^{-i \omega t} Y_{sm}(\theta,\psi),
\end{equation}
where $Y_{sm}$ are the spherical harmonics  ($s$ is the multipole number). 
When presenting and comparing our numerical results, we shall use units in which 
$r_{0} = 1$ and $c_{s} = 1.$ This can be achieved with the substitution 
$r\to r_{0} r$ and $\omega \to \omega c_{s}.$ Consequently, 
the radial equation assumes the general form~(\ref{master}) with
\begin{equation}
 V(r) = \left(1-\frac{1}{r^{4}} \right)\left(\frac{s(s+1)}{r^{2}} + \frac{4}{r^{6}} \right).
 \label{hy4}
\end{equation}

\section{$D$-dimensional Schwarzschild-Tangherlini 
black holes vs. acoustic black holes}
\label{sekcja3}

We start our discussion of the acoustic black holes with the observation that the
radial equations of the (2+1) and (3+1)-dimensional black holes are formally similar
to the analogous equations of $(D = 5)$-dimensional and $(D = 7)$-dimensional
Schwarzschild-Tangherlini black holes, respectively.  More precisely, the radial equations
describing perturbations of the acoustic black holes can be obtained from the master
equation of the $D$-dimensional Schwarzschild-Tangherlini black hole by the
appropriate choice of parameters.  Therefore, before addressing the particular case
of hydrodynamic black holes, let us briefly discuss the properties of the
$D$-dimensional solution that are important from our perspective.

\subsection{Master equation.}

The $D$-dimensional generalization of the Schwarzschild black hole is described by
the line element
\begin{equation}
 ds^{2} = -f(r) d r^{2} dt^{2} + \frac{d r^{2}}{f(r)} + r^{2} d\Omega^{2}_{D-2}
\end{equation}
where
\begin{equation}
 f(r) = 1 - \frac{\mathcal{M}}{r^{3-D}}
\end{equation}
and $d\Omega^{2}_{D-2} $ is a metric on a unit $(D-2)$-dimensional sphere. 
The black hole mass $M$ is related to the parameter $\mathcal{M}$ by a simple
relation
\begin{equation}
M = \frac{(D-2)\omega_{D-2} \mathcal{M}}{16 \pi G_{D}}
\end{equation}
with 
\begin{equation}
\omega_{D-2} = \frac{2 \pi^{(D-2)/2}} {\Gamma[(D-1)/2]}. \end{equation}
In what follows,  we shall use the  geometric units and express our numerical results 
assuming $\mathcal{M} = 1.$

Now, the
differential (master) equation describing the massless scalar, gravitational tensor,
gravitational vector,  electromagnetic vector, and electromagnetic scalar
perturbations of the $D$-dimensional Schwarzschild-Tangherlini black hole ($D>4$) can
be written in a compact form~\cite{Hartnoll,Kodama1,Kodama2}:
\begin{equation}
\frac {d^2} {d r_{\star}^{2}} \psi + \left\{\omega^{2}  - f(r) \left[ \frac {l(l+D-3)} {r^2}
+ \frac {(D-2)(D-4)} {4r^2} + \frac {(1-j^2)(D-2)^2} {4r^{D-1}} \right]\right\}\psi = 0,
\label{master1}
\end{equation}
where $j$ is given by
\begin{equation}
j=
\begin{cases}
0, &       \text{massless scalar and gravitational tensor perturbations}.  \\ 
2, &       \text{gravitational vector perturbations},\\
\frac{2}{D-2}, &   \text{electromagnetic vector  perturbations}\\
2 -\frac{2}{D-2}, & \text{electromagnetic scalar  perturbations}.\
\end{cases}
\label{typy_zab}
\end{equation}
Of course, from a general point of view, we are not confined to the values of the parameter
$j$ listed in Eq.~(\ref{typy_zab}). By doing so, we can gain a new interpretation of Eq.~(\ref{master1})
at the expense of losing its original interpretation. We will make use of this fact later in this work.

The quasinormal modes are defined as the solutions of the master equation that are
purely ingoing as $r_{\star} \to -\infty$ and purely outgoing as $r_{\star} \to
\infty$ with the asymptotic behaviors $e^{ -i \omega r_{\star}}$ and $e^{i \omega
r_{\star}},$ respectively.  Now, the standard procedure is to construct a series
expansion of the perturbation $\psi$ with the required asymptotics. The first choice
that has been adopted in this paper is to take
\begin{equation}
 \psi(r) = \left( \frac{r-1}{r} \right)^{-i\omega/(D-3)} e^{i \omega r } 
\sum_{k=0}^{\infty} a_{k} \left(\frac{r-1}{r} \right)^{k},
                     \label{expansion_even}
\end{equation}
for the even-dimensional case and
\begin{equation}
 \psi(r) = \left(\frac{r-1}{r+1}\right)^{-i \omega/(D-3)} e^{i \omega r} 
\sum_{k=0}^{\infty} a_{k} \left(\frac{r-1}{r} \right)^{k}.
                     \label{expansion_odd}
\end{equation}
for the odd-dimensional case.  
The even-dimensional expansion is given here for completeness.
Upon the  substitution of  (\ref{expansion_even}) and
(\ref{expansion_odd}) into the master equation one obtains the recurrence relations
for the expansion coefficients $a_{k}.$ As a result, we will obtain $(2 D-5)$-term
recurrence relation for the even-dimensional black hole and $(2 D - 6)$-term
recurrence for the odd-dimensional black hole.

It should be noted that, depending on a specific dimension, the functions
(\ref{expansion_even}) and (\ref{expansion_odd}) can be  modified to yield  even
simpler recurrences. Indeed, for the $(3+1)$-analogue black hole, the expansion
\begin{equation}
 \psi(r) = \left(\frac{r-1}{r+1}\right)^{-i \omega/4} e^{i \omega r} e^{-i \omega \arctan( r)/2}
 \sum_{k=0}^{\infty} a_{k} \left(\frac{r-1}{r} \right)^{k}
\end{equation}
leads to the six-term recurrence. It is an improvement over the method based on the
expansion (\ref{expansion_odd}).  All the calculations presented in this paper
regarding the $(3+1)$-dimensional hydrodynamic black hole have been carried out using
both expansions, leading to the results that are in perfect agreement with each
other.

\subsection{The recurrence relations for the  $(2+1)$-dimensional acoustic black hole }

The radial equation of the $(2+1)$-dimensional acoustic black hole is formally identical
to the analogous equation describing the electromagnetic vector perturbations of the 
five-dimensional Schwarzschild -Tangherlini black hole.
Indeed, putting $D = 5,$ $j = 2/3$, and $l = m - 1$ for $m\geq 1,$ 
the master equation (\ref{master1}) reduces
to Eq.~(\ref{master}) with the potential term given by (\ref{hy3}), and one can use
(\ref{eq_a1}) and (\ref{eq_a2}) to obtain the four-term recurrence relation
\begin{equation}
 \alpha_{k} a_{k+1} + \beta_{k} a_{k} + \gamma_{k} a_{k-1} + \delta_{k} a_{k-2} =0,
                    \label{recurrence_1}
\end{equation}
where
\begin{equation}
\begin{array}{lcl}
\alpha_{k} &=& -8 (1+k)(1+k -2 \rho),\\
\beta_{k} &=& 20k^{2} + 4 m^{2} + 4 k (5-16\rho) + 4 (1-4 \rho)^{2},\\
\gamma_{k} &=& 6 -16 k^{2} + 32 k \rho,\\
\delta_{k} &=& -3 - 4 k + 4 k^{2}.
 \end{array}
 \label{rec1}
\end{equation}
Here  $a_{-2} = a_{-1} =0,$
and (following tradition) we introduced $\rho = i \omega/2.$ 
Of course, one can start with Eqs.~(\ref{master}) and
(\ref{hy3}) and construct the solution directly, identifying the singular points of
the differential equations. Then, it suffices to substitute expansion
(\ref{expansion_odd}) into the equation (\ref{master}) and use the standard
techniques to construct the coefficients of the recurrence relation.

\subsection{The recurrence relations for the  (3+1)-dimensional acoustic black hole }

A simple analysis shows that, formally, the radial equation of the
$(3+1)$-dimensional acoustic black hole is a special case of the $D$-dimensional master equation
with $D = 7,$ $j = 3/5$ and $l = (2 s-3)/2.$  
Unlike the previous case, however, this type of perturbation is not listed in Eq.~(\ref{typy_zab}), 
and the multipole number $l$ is fractional. Now, to construct the recurrences, 
it suffices to make use of either (\ref{eq_a3}) or (\ref{eq_a4}). Alternatively,
one may start with the potential (\ref{hy4}) from the very beginning.
\subsubsection{The eight-term recurrence relations}

Inserting the series expansion (\ref{expansion_odd}) into the master equation
one obtains the eight-term recurrence relations
\begin{equation}
 a_{k+1}\, \alpha_{k} + a_{k} \,\beta_{k} + a_{k-1}\,\gamma_{k}
+ a_{k-2} \,\delta_{k} + a_{k-3}\, \kappa_{k} + a_{k-4}\, \lambda_{k}+
 a_{k-5} \,\mu_{k} + a_{k-6}\, \nu_{k} = 0,
\end{equation}
with
\begin{equation}
 a_{-6} = a_{-5} = a_{-4} = a_{-3} = a_{-2} = a_{-1} = 0,
\end{equation}
where
\begin{equation}
\begin{array}{lcl}
\alpha_{k} &=& -8 (k+1) (k-\rho +1) ,\\
\beta_{k} &=& 2 \left(18 k^2+k (14-32 \rho )+14 \rho ^2-14 \rho +s(s+1) + 4\right),\\
\gamma_{k} &=&  -2 \left(36 k^2-2 k (34 \rho +9)+33 \rho ^2+16 \rho +s(s+1) + 2\right),\\
\delta_{k} &=& 84 k^2-152 k (\rho +1)+69 \rho ^2+142 \rho +s(s+1) + 52,\\
\kappa_{k} &=& -2 \left(31 k^2-k (49 \rho +97)+18 \rho ^2+80 \rho +60\right),\\
\lambda_{k} &=& 29 k^2-3 k (12 \rho +43)+9 \rho ^2+84 \rho +116 ,\\
\mu_{k} &=&-8 k^2+k (6 \rho +46)-18 (\rho +3) ,\\
\nu_{k} &=& k^2-7 k+10.
 \end{array}
 \label{rec2}
 \end{equation}

\subsubsection{A different choice of the radial function $\psi(r)$}

Now, let us consider a slightly modified radial function $\psi(r)$ defined as
\begin{equation}
 \psi(r) = \left(\frac{r-1}{r+1}\right)^{-i \omega/4} e^{i \omega r} 
e^{-i \omega \arctan( r)/2} \sum_{k=0}^{\infty} a_{k} \left(\frac{r-1}{r} \right)^{k}.
\end{equation}
It can be shown that such a choice leads to the six-term recurrence
\begin{equation}
 a_{k+1}\, \alpha_{k} + a_{k} \,\beta_{k} + a_{k-1}\,
\gamma_{k}+ a_{k-2} \,\delta_{k} + a_{k-3}\, \kappa_{k} + a_{k-4}\, \lambda_{k} = 0,
\end{equation}
with $  a_{-4} = a_{-3} = a_{-2} = a_{-1} = 0,$
where
\begin{equation}
\begin{array}{lcl}
\alpha_{k} &=& -4 (k+1) (k-\rho +1), \\
\beta_{k} &=&  4 + 14 k^{2} + s(s+1) + 2 k (7 -12 \rho) + 12 \rho (\rho -1), \\
\gamma_{k} &=& -20 k^2+28 k \rho -8 \rho ^2+4, \\
\delta_{k} &=& 15 k^2-k (16 \rho +15)+4 \rho ^2+8 \rho -6, \\
\kappa_{k} &=& -6 k^2+4 k (\rho +3)-4 \rho +2, \\
\lambda_{k} &=&  (k-3) k .\\
\end{array}
\label{rec3}
\end{equation}
In what follows, we shall use both recurrence relations.

\section{The methods based on the recurrence relations}
\label{sekcja4}
\subsection{The Hill determinant method}

The recursion coefficients determined in the previous section can be utilized in
either the Hill determinant method or the continued fraction method.  In the context
of the quasinormal modes of the black holes, the Hill determinant method has been used
for the first time by Majumdar and Panchapakesan in Ref.~\cite{Pancha}.  For years it
has been treated as a lesser relative to its more prominent cousins, as for example
the Leaver method. The Hill determinant method has been criticized for rapid
deterioration of the accuracy of the complex frequencies of the quasinormal modes,
especially for the overtones, and the lack of a natural estimator of the error
caused by  a finite dimension of the Hill matrices.  While some of the critique may be
justified, it does not mean that the method should be abandoned. Even in its simplest
version, the approximate results may be used as initial values for more sophisticated
methods.  Further improvements may lead to a substantial increase of its accuracy,
which, with some additional effort, can be better than the accuracy of the Leaver
method in its original form.  The Hill determinant method is one of the three methods
used in this paper. Here we will present only its main features.

We start with the four-term recurrence. The condition that the nontrivial solutions 
of the recurrence relation (\ref{recurrence_1}) exist is given by  
\begin{equation}
 \det \mathcal{H} = 0,
 \label{dHill}
\end{equation}
where $\mathcal{H}$ is the banded Hill matrix of width 4
\begin{equation} \mathcal{H} = 
\begin{bmatrix}
 \beta_{0}  & \alpha_{0}  &              &              &              &            &              &                &              &               &              &   \\ 
 \gamma_{1} & \beta_{1}   & \alpha_{1}   &              &              &            &              &                &              &               &              &   \\
 \delta_{2} & \gamma_{2}  & \beta_{2}    & \alpha_{2}   &              &            &              &                &              &               &              &   \\
            & \delta_{3}  &  \gamma_{3}  & \beta_{3}    &   \alpha_{3} &            &              &                &              &               &              &   \\
            &             &              &              &              & \ddots \ddots    &              &                &              &               &              &   \\ 
            &             &              &              &              &            & \delta_{n-1} & \gamma_{n-1}   & \beta_{n-1}  & \alpha_{n-1}  &              &   \\
            &             &              &              &              &            &              & \delta_{n}     & \gamma_{n}   & \beta_{n}     & \alpha_{n}   &   \\
            &             &              &              &              &            &              &                &              &               &              &   \ddots \ddots
\end{bmatrix}
\end{equation}
Consider the determinants of the $n \times n$ submatrices $\mathcal{H}_{n}$ with
the main diagonal $\beta_{0},...,\beta_{n-1},$ i.e., the leading principal minors.  
The resulting equations are polynomials and can be solved order by order. 
Since the modern procedures  for
calculating the determinants of the matrices are very robust and the Hill matrices
are sparse there is no need to use Gauss elimination to transform the recurrence
relations to the tridiagonal case.  Alternatively, we can use a simple formula for
constructing the determinants. Indeed, denoting by $h_{n}$ the determinant of
$(n+1)\times(n+1)$    matrix, for $n \geq 3$ one has
\begin{equation}
 h_{n} = \beta_{n} h_{n-1} -\gamma_{n} \alpha_{n-1} h_{n-2} + \delta_{n} \alpha_{n-1} \alpha_{n-2} h_{n-3}.
 \label{rec_hill_1}
\end{equation}

On the other hand, the general procedure in its original form is as follows: First, we transform the Hill matrix
to the tridiagonal one
\begin{equation} \mathcal{H'} =
\begin{bmatrix}
 \beta'_{0} & \alpha'_{0} &              &            &             &                &             &                &              &\\
 \gamma'_{1}& \beta'_{1}  & \alpha'_{1}  &            &             &                &             &                &              &\\
            & \gamma'_{2} & \beta'_{2}   & \alpha'_{2}&             &                &             &                &              &\\
            &             &              &            &   \ddots \ddots   &                &             &                &              &\\
            &             &              &            &             & \gamma'_{n-1}  & \beta'_{n-1}& \alpha'_{n-1}  &              &\\
            &             &              &            &             &                & \gamma'_{n} & \beta'_{n}     & \alpha'_{n}  &\\
            &             &              &            &             &                &             &                &              & \ddots  \ddots
\end{bmatrix}
\end{equation}
and  construct the determinant of the $N\times N$ Hill matrix for a sufficiently big $N$.
Denoting once again the determinant of $(n+1)\times (n+1)$ matrix by $h'_{n},$
one arrives at the simple relation
\begin{equation}
 h'_{n} = \beta'_{n} h'_{n-1} -\gamma'_{n} \alpha'_{n-1} h'_{n-2},
\end{equation}
which can formally be obtained  from (\ref{rec_hill_1}) by putting $\delta_{n} =0.$
The resulting equations are functions of $\rho$ and can be solved numerically.
Further, the boundary and the stability conditions select on the complex plane some
subset of the solutions with negative imaginary part of the frequency, which are
identified as the quasinormal modes~\cite{Pancha}.

In a modified approach~\cite{ja_dim5}, we calculate solutions of the determinant
equation order by order for increasing $k$ up to some $N$ and look for stable roots.
The stable roots that approximate the exact quasinormal frequency migrate on the
complex plane but they remain in the basin of convergence.  We consider the roots as
stable if for increasing $k$ their location does not change to the assumed precision.
Unfortunately, in many cases, such as when calculating higher overtones or when
greater precision of the results is required, the convergence of the approximants may be
slow.  In this case one can increase $N$ and accelerate convergence of the
approximants using some standard techniques, as for example, the Pad\'e summation or
the Wynn algorithm. 

The $(3+1)$ hydrodynamic black hole can be can be treated similarly. First, we
construct the banded $(k\times k)$ matrices $(k = k_{0},...,N)$ of width eight (or
six) and calculate their determinants.  Subsequently we solve the resulting equations
and search for stable roots.  Finally, we accelerate convergence of the approximants
$\omega_{k}.$ Since the construction of the tridiagonal matrices requires the
execution of 5 (or 3) consecutive Gaussian eliminations, we will not employ this
approach here.

\subsection{The continued fractions method}

The method of continued fractions is widely used in the context of the quasinormal
modes. It is based on the profound relationship between the continued fractions and
the three-term recursions.  Here we give only a few basic facts that will be needed 
in our discussion. 

Let us consider a four-term recurrence of the type (\ref{recurrence_1}). Generalizations to
more complicated cases are obvious. Suppose that it has been transformed to a three-term recurrence
\begin{equation} 
  \alpha'_{k} a_{k+1} + \beta'_{k} a_{k} + \gamma'_{k} a_{k-1} =  0
\end{equation}
(with $a_{-1} =0$)
by the appropriate Gauss elimination. 
The primed coefficients are given by
\begin{equation}
 \alpha'_{k} = \alpha_{k},
\end{equation}
\begin{equation}
  \beta'_{k} = \begin{cases}
 \beta_{k}, & \, {\rm for}\,\, k = 0,1 \\
 \beta_{k} - \delta_{k} \alpha'_{k-1}/\gamma'_{k - 1}  &\, {\rm for}\, \,k \geq 2
 \end{cases}
\end{equation}
and 
\begin{equation}
  \gamma'_{k} = \begin{cases}
 \gamma_{k}, & \, {\rm for}\,\, k = 0,1 \\
 \gamma_{k} - \delta_{k} \beta'_{k-1}/\gamma'_{k - 1}  &\, {\rm for}\, \,k \geq 2.
 \end{cases}
\end{equation}
As is well-known, the convergence condition for the series
expansion (which is simultaneously the condition for the quasinormal modes) can be
expressed in the form of an infinite continued fraction:
\begin{equation}
 \beta'_{0} - \dfrac{\alpha'_{0} \gamma'_{1}}
                    {\beta'_{1} - \dfrac{\alpha'_{1} \gamma'_{2}}
                                       {\beta'_{2} - \dfrac{\alpha'_{2} \gamma'_{3}}{\beta'_{3} - \dots}}} = 0
                                       \label{cont1}
\end{equation}
or, equivalently, in the more popular notation,
\begin{equation}
 \beta'_{0} - \frac{\alpha'_{0} \,\gamma'_{1}}{\beta'_{1} -}\, \frac{\alpha'_{1}\,
 \gamma'_{2}}{\beta'_{2} -}\, \frac{\alpha'_{2}\, \gamma'_{3}}{\beta'_{3} -}\dots = 0.
 \label{cont2}
\end{equation}
Inverting Eq.~(\ref{cont1}) $n$ times, one obtains
\begin{equation}
 \beta'_{n} -\frac{\alpha'_{n-1} \,\gamma'_{n}}{\beta'_{n-1} -}\,\frac{\alpha'_{n-2} \,
 \gamma'_{n-1}}{\beta'_{n-2} -}\,\dots -\frac{\alpha'_{0} \gamma'_{1}}{\beta'_{0}} =
 \frac{\alpha'_{n} \,\gamma'_{n+1}}{\beta'_{n+1} -}\, \frac{\alpha'_{n+1}\,
 \gamma'_{n+2}}{\beta'_{n+2} -}\, \frac{\alpha'_{n+2}\, \gamma'_{n+3}}{\beta'_{n+3} -}\dots.
 \label{cont3}
\end{equation}
If the recurrence is constructed for the radial equation (\ref{master}), its
minimal solutions correspond to the quasinormal modes. Now, our strategy is as follows:
First, we (numerically) solve Eq.~(\ref{cont1}) truncated at some index $N$ and identify
the (approximate) quasinormal frequencies using some low-cost method. Subsequently, we increase
$N$ and analyze the migration of the quasinormal modes on the complex plane. In doing so,
we obtain a series $\omega(N)$ that may or may not converge rapidly to the limiting value.
In the latter case, we accelerate the convergence of the series.

\section{The WKB-based methods}
\label{sekcja5}

The WKB-based methods are among the most popular approaches for calculating the
complex frequencies of the quasinormal modes. This is because the calculations  are
relatively simple and in many cases yield quite accurate results. In this approach,
we attempt to construct the approximate solutions of Eq.~(\ref{master}) using 
a modification of the WKB method. The asymptotic conditions near the event horizon
and the conditions for the quasinormal modes lead to the equation relating 
frequencies and the behavior of the potential term near its maximum. Denoting $Q(x) = \omega^{2} -V(x)$
and its consecutive derivatives calculated  at by $x_{0}$ by $Q_{0},\, Q_{0}', \,Q_{0}'',...,$
this equation can be written in the form
\begin{equation}
 \frac{ i Q_{0}}   {\varepsilon \sqrt{  2Q''_{0}}} -\sum_{k=2}^{N} 
 \varepsilon ^{k-1} \Lambda_{k}= n+ \frac{1}{2}.
 \label{master2}
\end{equation}
Here, each $\Lambda_{k}$ is constructed from the derivatives of $Q(x)$ 
calculated at $x=x_{0},$ where the potential attains its maximum, and
$\varepsilon$ is the expansion parameter.
Solving (\ref{master2}) with respect to $\omega$ one gets 
\begin{equation}
 \omega^{2} = V(x_{0}) - i \left( n + \frac{1}{2} \right) \sqrt{2 Q''_{0}} \varepsilon
 -i  \sqrt{2 Q''_{0}} \sum_{k = 2}^{N} \varepsilon^{k} \Lambda_{k}.
                  \label{main1}
\end{equation}

The above formula can be used to perform a brief historical survey of the simple
WKB-based methods.  Retaining the first two terms in Eq.~(\ref{main1}) results in the
Schutz-Will approximation~\cite{wkb0} (see also \cite{Bahram}). Furthermore,
including the two next terms of the expansion, $\Lambda_{2}$ and $\Lambda_{3},$ yields
the well-known and widely used Iyer-Will method~\cite{wkb1}. Finally, Konoplya
calculated the terms $\Lambda_{k}$ up to $N = 6$ and demonstrated that in many cases
the results are close to the results of the exact numerical
calculations~\cite{Konoplya6}. In all this methods the $\Lambda_{k}$ parameters are
just summed and since $|\Lambda_{k}|$ grow fast with $k$ one may ask if there is any
point in increasing the number of retained $\Lambda$-terms. The answer to this doubts has
been given in Ref.~\cite{jaOP}, in which all the terms up to $N=13$ has been
calculated analytically and since the summation of the $\Lambda_{k}$ terms is
generally a bad strategy in this regard, it has been proposed to construct the Pad\'e
approximants of Eq.~(\ref{main1}).  Specifically, in~\cite{jaOP} it has been shown
that for the well-documented complex frequencies of the quasinormal modes of the
Schwarzschild and the Reissner-Nordstr\"om black holes, the Pad\'e summation always
gives the results that are in an excellent agreement with the exact numerical
calculations.  The methods presented in Ref.~\cite{jaOP} has been subsequently
extended in Ref.~\cite{jago}, where the analytical calculations has been carried out
up to $N =16.$ Unfortunately, the analytic calculations of the higher-order terms are both
time consuming and pose high demands on the computer resources, as the complexity of 
the $\Lambda$-terms rapidly increase with $k.$  Since the dynamics
of the calculations suggests that the most interesting are the higher orders of the WKB
approximation it is natural that to make real progress and to make  the
calculations tractable one has to switch to numerics. The strategy would be,
therefore, as follows. First, we calculate numerically all $\Lambda_{k}$ up to
required order and subsequently we accelerate the convergence of the sequence/series
(or even make them convergent) using some standard methods, as for example, the
Pad\'e summation~\cite{jaOP}, the $\epsilon$-Wynn acceleration~\cite{jago}, the Borel
summation~\cite{hatsuda}, and the Borel-Le Roy summation~\cite{ja_dim5}. Finally, if
possible, we compare the thus obtained results with the results constructed within
the framework of other methods, as for example the Hill determinant method and its
modifications~\cite{Pancha} or the method of continued fraction~\cite{Leaver}.

In a very interesting development~\cite{OlegZ}, Zaslavskii demonstrated that the problem 
of calculating complex frequencies of quasinormal modes can be reformulated as 
a bound-state problem. Similar
considerations have been presented in Refs.~\cite{Blome,val1}. Specifically,
Zaslavskii demonstrated that it is possible to reconstruct the main result of the
Iyer and Will paper studying an anharmonic oscillator with corrections up to
sixth-order. It should be noted that Zaslavskii's method may not be optimal for
constructing the higher-order terms of the WKB approximation.  Fortunately, in  a
very interesting paper~\cite{Sulejman} Sulejmanpasic and \"Unsal  extended to the
arbitrary polynomial corrections the classical results of Bender and Wu~\cite{BW}.
Their paper is accompanied with a powerful computer algebra package for fast and
reliable calculations of the really high orders of the perturbation expansion. This
package has been used in very accurate calculations of the complex frequencies of
the  quasinormal modes in Refs.~\cite{hatsuda,jago,ja2020b,ja_dim5,kisel}.

\section{Quasinormal modes}
\label{sekcja6}

\subsection{$(2+1)$-dimensional analogue black hole}

In determining the frequencies of the quasinormal modes of the $(2+1)$-dimensional black holes, we
restricted ourselves to the modes satisfying relations $1 \leq m \leq 6$ and $0 \leq
n \leq m + 1.$ The reason for this choice is twofold. First, the complex
frequencies of the overtones (labeled here by $n$) of the low-lying fundamental
modes (characterized by the azimuthal number $m$) are hard to be calculated precisely
within the framework of the WKB-based method.  Secondly, the low-lying modes are more
likely to be detectable in laboratory settings, as the higher overtones are damped
quickly and virtually play no role.  Since our main task is to provide highly
accurate results, that are correct to, say, 9 decimal places we decided to accept the
result only if for a given $m$ and $n$ there is an agreement between the results
obtained within the framework of at least two independent methods. In the vast
majority of cases, we compared  the frequencies obtained  using the WKB method and
the  Hill determinant method. The continued fraction method has been used to check
the correctness of the  result to 9 decimal places.  The computational strategy is as
follows. First we calculate the frequencies of the quasinormal modes using either the
higher-order WKB-Pad\'e method or the higher-order WKB with the Wynn convergence
acceleration algorithm. We deem the result acceptable if the frequencies remain
unchanged to the prescribed accuracy as the order of the WKB increases. Typically, to
ensure high accuracy for the overtones of the low-lying modes it is necessary to
conduct the calculations up to some really big $N$ and to retain as many digits in
the decimal representation as needed.  The final results were rounded to 64 decimal
places. Subsequently, the frequencies thus calculated, when rounded to two decimal
places, were used as the starting values in the Hill determinant method. For a given
$m$ and low $n$, the stable roots of the determinant equations, as functions of the
dimension of the Hill submatrices, rapidly approach the exact frequency and one can easily
obtain reasonable answers without additional effort. Unfortunately, the accuracy
deteriorates as the overtone number increases, and one is forced to use the convergence
acceleration techniques.  Here, regardless of the type of the mode,  we used the Wynn algorithm
to accelerate the convergence of the approximants. This approach makes it easier 
to automate the calculations.

\begin{table}
	\caption{Quasinormal modes of the $(2+1)$-dimensional hydrodynamic black
hole.  $m$ is the azimuthal number, $n$ the overtone number, $N_{max}$ is the maximal
dimension of the Hill matrix, $\omega$ is the complex frequency calculated within the
frameworks of the Hill determinant method and the continued fraction method and
$\Delta \omega = \omega_{Hill} - \omega_{WKB}$. The high precision results of
$\omega$ are rounded to 9 decimal places.}
\begin{ruledtabular}
\begin{tabular}{ccccc}
$ m$ & $n$ & $N_{max}$ & $\omega$  &$ \Delta \omega = \omega_{Hill} - \omega_{WKB}$  \\
\colrule
$1$&$  0$&$  300$&$ 0.4068326197 - 0.3412361181 i$&$ -5.16\times10^{-14} + 1.29\times10^{-14} i$\\ 
$ $&$1 $&$  600$&$ 0.1974857187 - 1.2327917501 i$&$  1.82\times10^{-6}  - 4.5\times10^{-7} i$\\ 
$ $& $  2$&$  900$&$ 0.091780332  - 2.246129421  i$&$  0.001055 - 0.000364 i$\\ 
$2$&$  0$&$  300$&$ 0.9527280877 - 0.3507394957 i$&$ -4.17\times10^{-25} + 2.16\times10^{-25} i$\\ 
$ $&$  1$&$  600$&$ 0.7855826085 - 1.1248440726 i$&$ -3.02\times10^{-21} - 6.4\times10^{-22} i$\\  
$ $&$  2$&$  900$&$ 0.582309540  - 2.062341146 i$&$  -1.37\times10^{-15} + 5.25\times10^{-15} i$\\ 
$ $&$  3$&$ 1200$&$ 0.451386741 -  3.078412997 i$&$  2.02\times10^{-11} + 1.56\times10^{-11} i$\\ 
$3$&$  0$&$ 100$&$  1.468540697 - 0.352425533 i$&$ -6.32\times10^{-19} - 4.49\times10^{-19} i$\\ 
$ $&$  1$&$  150$&$ 1.348492532 - 1.089552723 i$&$ -1.42\times10^{-16} + 9.5\times10^{-17} i$\\ 
$ $&$  2$&$ 200$&$ 1.145380889 - 1.922656549 i$&$  4.51\times10^{-13} - 3.2\times10^{-14} i$\\ 
$ $&$  3$&$ 250$&$   0.942036373 - 2.870245548 i$&$ -6.6\times10^{-11} + 4.44\times10^{-10} i$\\ 
$ $&$  4$&$  300$&$ 0.790721695 - 3.880709660 i$&$ 2.36\times10^{-8} - 3.34\times10^{-8} i$\\ 
$4$&$ 0$&$   100$&$ 1.976452714 - 0.352959421 i$&$ 3.43\times10^{-22} + 8.1\times10^{-23} i$\\ 
$ $&$   1$&$ 150$&$ 1.884585741 - 1.076850329 i$&$ -8.1\times10^{-22} -    3.63\times10^{-21} i$\\ 
$ $&$ 2$&$ 200$&$ 1.714004250 - 1.856157434 i$&$   8.18\times10^{-19} - 8.32\times10^{-19} i$\\ 
$ $&$ 3$&$ 250$&$   1.501085485 - 2.721858168 i$&$ -9.65\times10^{-16} - 9.64\times10^{-16} i$\\ 
$ $&$ 4$&$   300$&$ 1.297445508 - 3.672813327 i$&$ 5.04\times10^{-13} - 5.45\times10^{-13} i$\\ 
$ $&$   5$&$ 350$&$ 1.133188616 - 4.676212409 i$&$ -5.33\times10^{-11} +   5.99\times10^{-11} i$\\ 
$5$&$ 0$&$ 100$&$   2.481187498 - 0.353188338 i$&$ -3.23\times10^{-25} + 3.85\times10^{-25} i$\\ 
$ $&$ 1$&$ 150$&$ 2.407098481 - 1.070993110 i$&$ 4.64\times10^{-25} + 8.1\times10^{-26} i$\\ 
$ $&$   2$&$ 200$&$ 2.264886375 - 1.824027440 i$&$ 1.67\times10^{-23} - 1.6\times10^{-24} i$\\ 
$ $&$ 3$&$ 250$&$   2.070976107 - 2.635991315 i$&$ -3.77\times10^{-21} + 4.17\times10^{-21} i$\\ 
$ $&$ 4$&$  300$&$ 1.855346091 - 3.521329941 i$&$ 9.0\times10^{-19} + 3.73\times10^{-18} i$\\ 
$ $&$ 5$&$ 350$&$ 1.651546102 - 4.473743454 i$&$ -1.31\times10^{-15} +  5.9\times10^{-16} i$\\ 
$ $&$ 6$&$ 400$&$ 1.478994088 - 5.470943809 i$&$   1.14\times10^{-13} - 1.17\times10^{-13} i$\\ 
$6$&$ 0$&$ 100$&$   2.984336621 - 0.353306306 i$&$ -1.27\times10^{-27} - 1.31\times10^{-27} i$\\ 
$ $&$ 1$&$   150$&$ 2.922346372 - 1.067823824 i$&$ -7.4\times10^{-29} + 1.79\times10^{-28} i$\\ 
$ $&$   2$&$ 200$&$ 2.801558110 - 1.806593575 i$&$ 4.53\times10^{-28} + 8.87\times10^{-28} i$\\ 
$ $&$ 3$&$ 250$&$ 2.630357486 - 2.586630216 i$&$ -3.69\times10^{-26} + 5.97\times10^{-26} i$\\ 
$ $&$ 4$&$   300$&$ 2.425007511 - 3.422706779 i$&$ 1.53\times10^{-23} - 1.61\times10^{-23} i$\\ 
$ $&$ 5$&$ 350$&$ 2.208931018 - 4.320895003 i$&$ 1.65\times10^{-21} - 1.028\times10^{-20} i$\\ 
$ $&$ 6$&$ 400$&$ 2.005051173 - 5.274064480 i$&$ 2.97\times10^{-18} - 3.7\times10^{-19} i$\\ 
$ $&$ 7$&$ 450$&$ 1.826891590 - 6.266212422 i$&$ -2.49\times10^{-16} + 2.40\times10^{-16} i$\\
\end{tabular}
\label{tab1}
\end{ruledtabular}
\end{table}

\begin{figure}
\centering
\includegraphics[width=12cm]{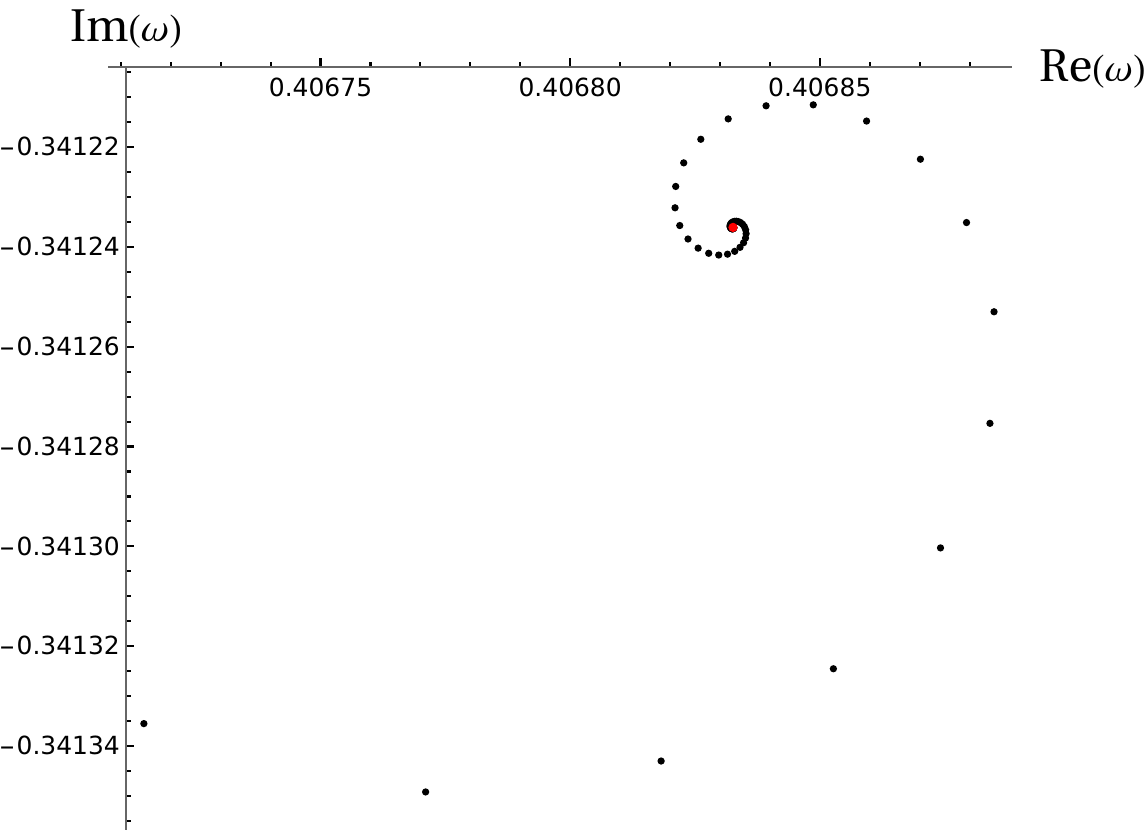}
\caption{The $(2+1)$-dimensional hydrodynamic black hole. Migration of the
approximants of the complex frequency of the $m=1$, $n = 0$ mode on the complex
plane. The red dot represents the limiting value calculated using the Wynn
acceleration algorithm.}
\label{fig0}
\end{figure}

\begin{figure}
\centering
\includegraphics[width=12cm]{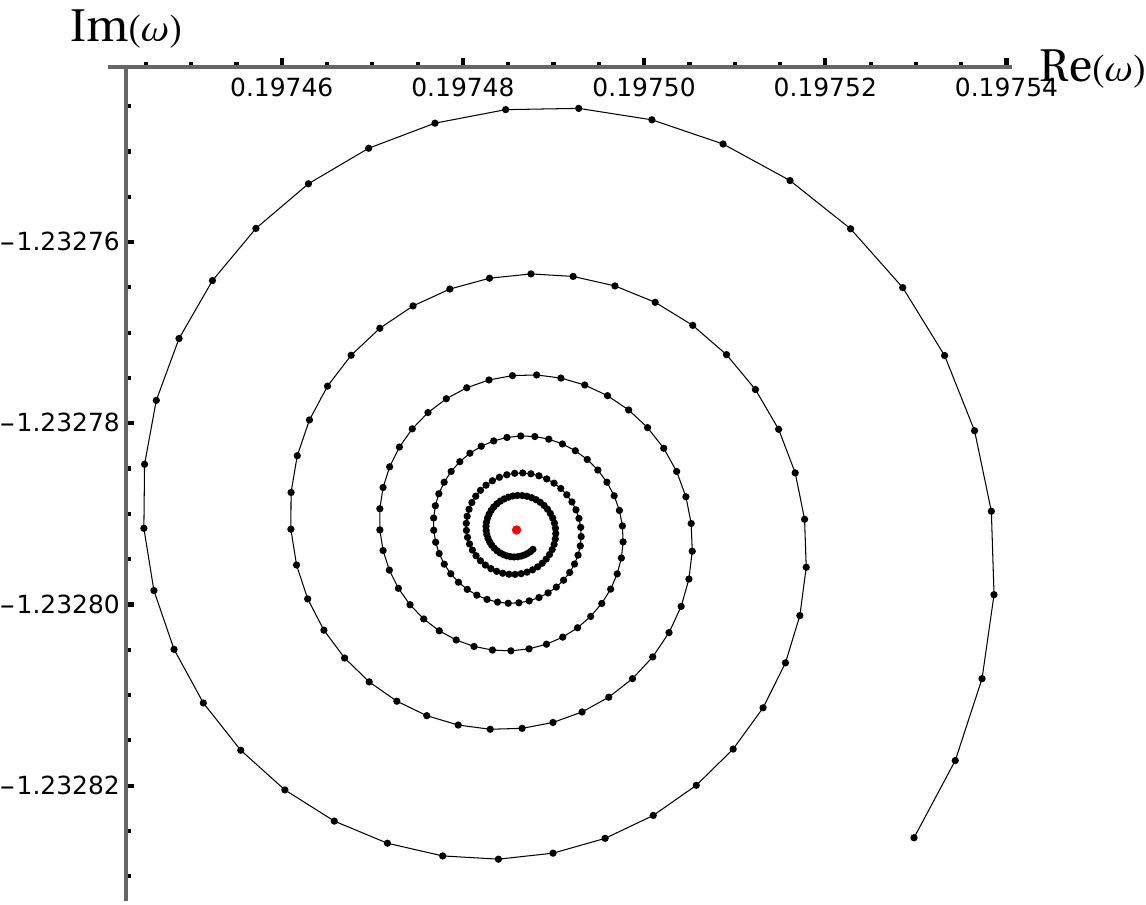}
\caption{The $(2+1)$-dimensional hydrodynamic black hole. Migration of the
approximants of the complex frequency of the $m=1$, $n = 1$ mode on the complex
plane. The red dot represents the limiting value calculated using the Wynn
acceleration algorithm.}
\label{fig1}
\end{figure}

\begin{figure}
\centering
\includegraphics[width=12cm]{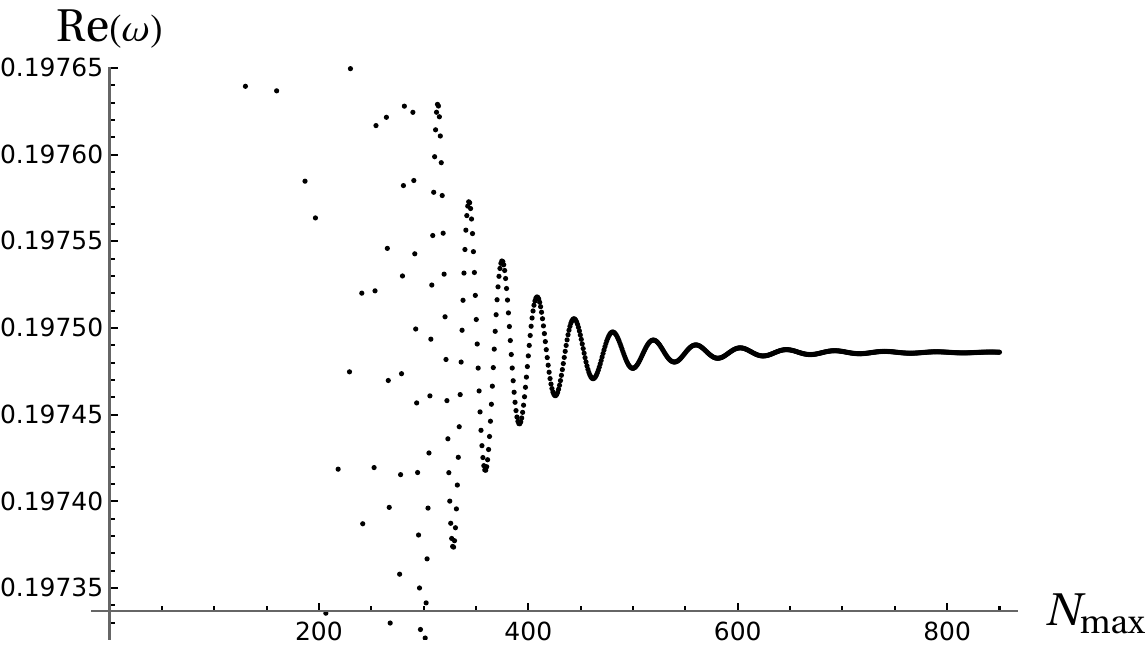}
\caption{The $(2+1)$-dimensional hydrodynamic black hole.  The real part of the
quasinormal frequency of the $m=1$, $n =1$ perturbation as a function of the
dimension of the Hill matrix.}
\label{fig2}
\end{figure}

\begin{figure}
\centering
\includegraphics[width=12.5cm]{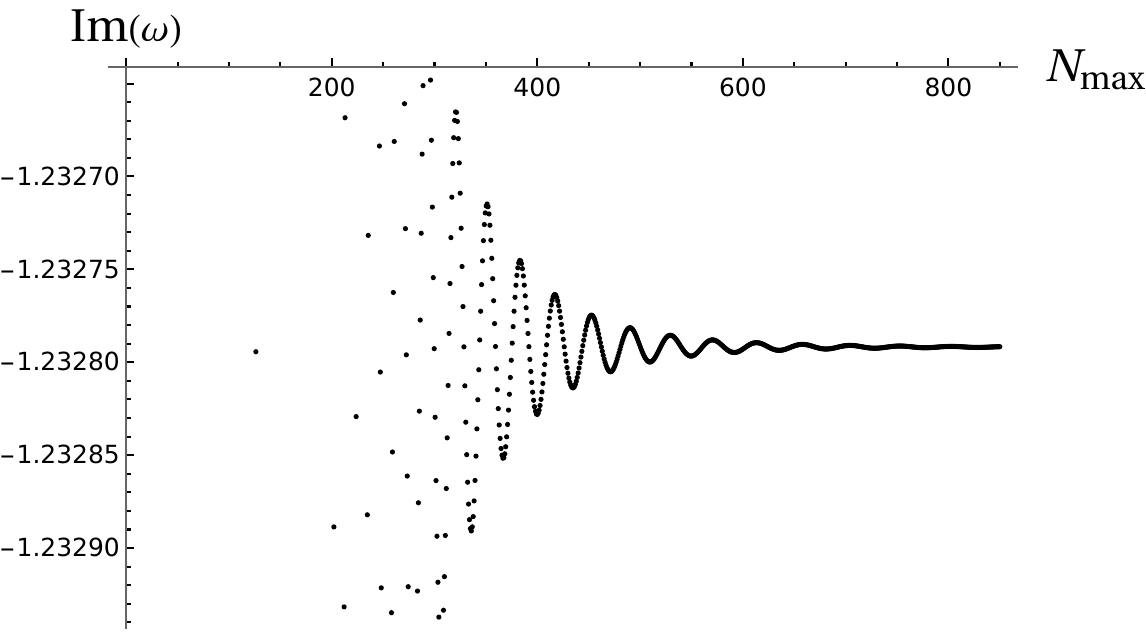}
\caption{The $(2+1)$-dimensional hydrodynamic black hole.  The imaginary part of the
quasinormal frequency of the $m=1$, $n =1$ perturbation as a function of the
dimension of the Hill matrix.}
\label{fig3}
\end{figure}

\begin{figure}
\centering
\includegraphics[width=12.5cm]{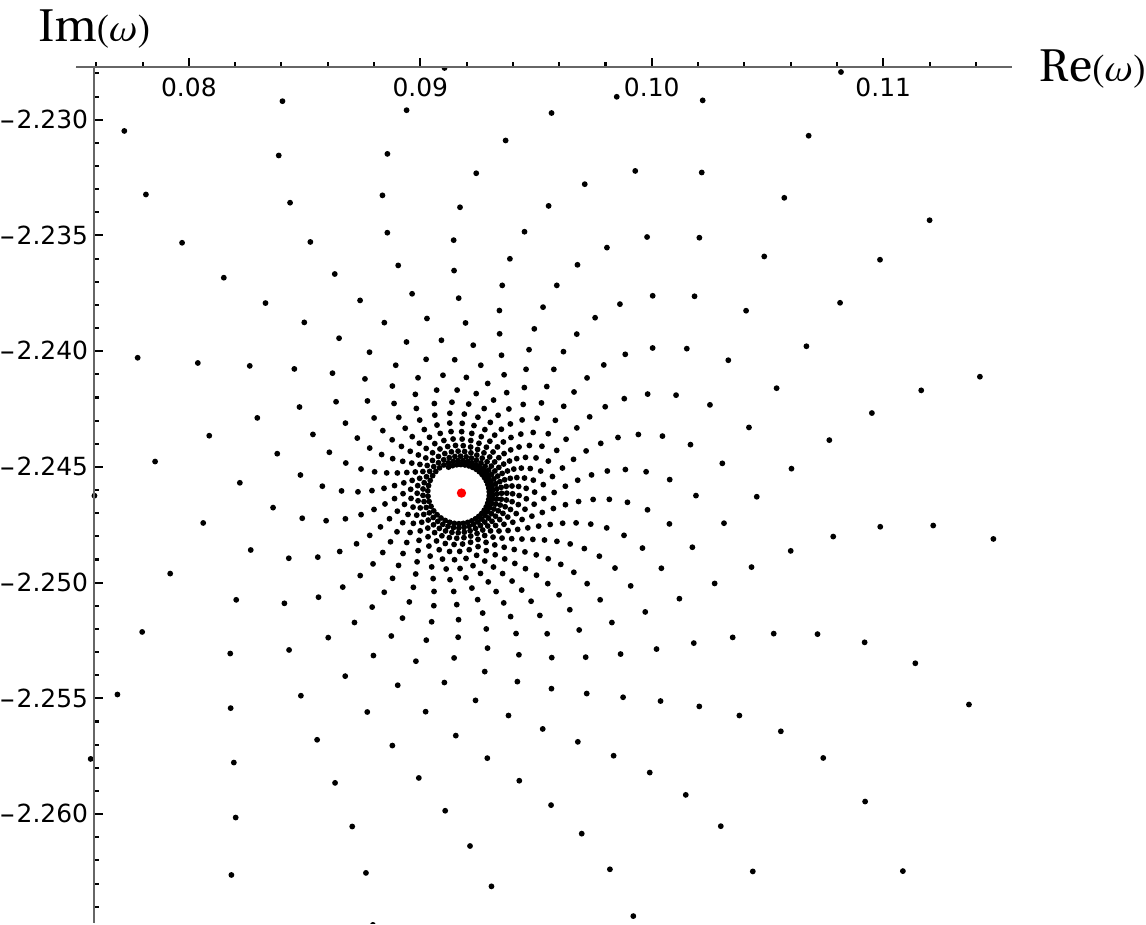}
\caption{The $(2+1)$-dimensional hydrodynamic black hole. Migration of the
approximants of the complex frequency of the $m=1$, $n = 2$ mode on the complex
plane. The red dot represents the limiting value calculated using the Wynn
acceleration algorithm.}
\label{fig4}
\end{figure}

\begin{figure}
\centering
\includegraphics[width=12.5cm]{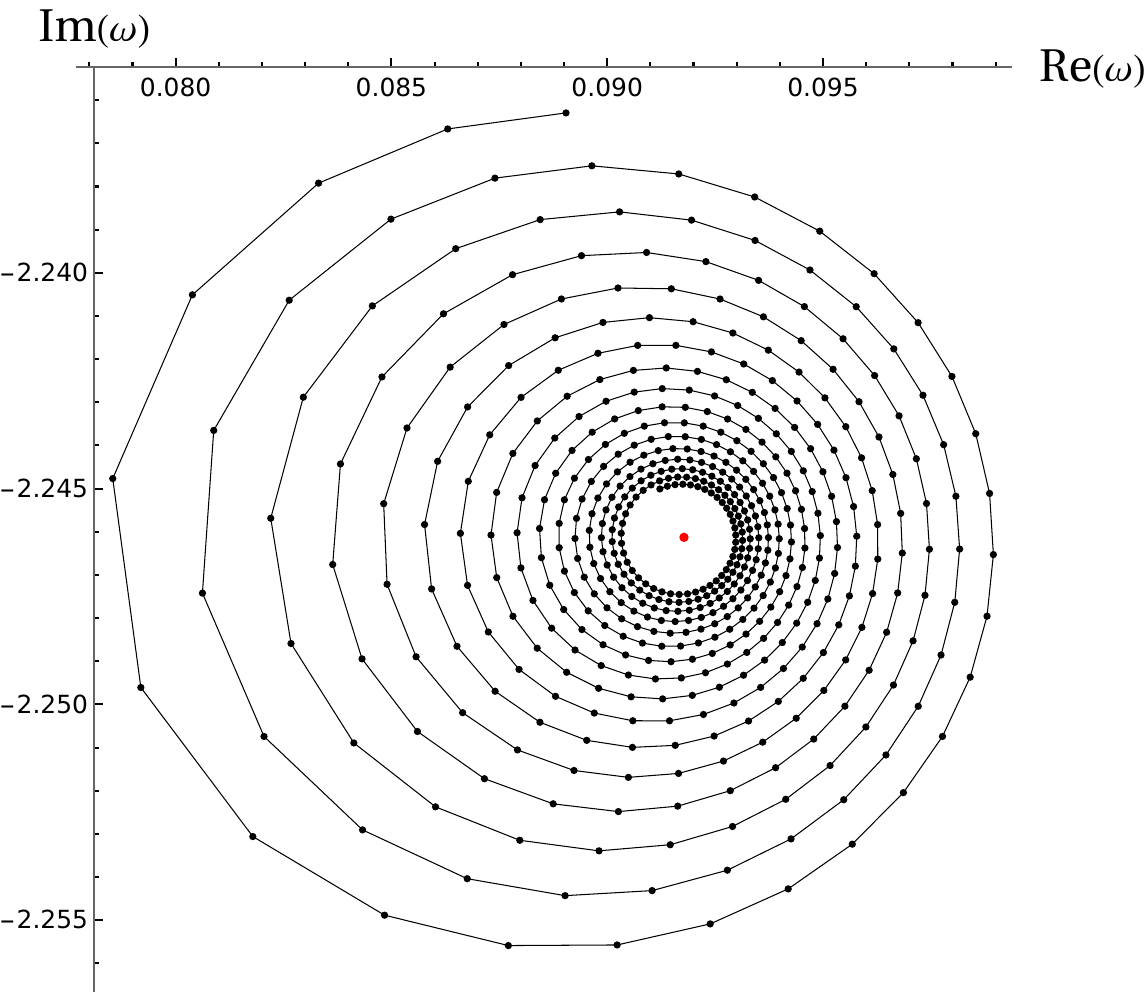}
\caption{The $(2+1)$-dimensional hydrodynamic black hole. Migration of the
approximants of the complex frequency of the $m=1$, $n = 2$ mode on the complex
plane. The red dot represents the limiting value calculated using the Wynn
acceleration algorithm. The $N$-th approximants in connected by a line segment with
the $N+1$.}
\label{fig5}
\end{figure}

\begin{figure}
\centering
\includegraphics[width=12.5cm]{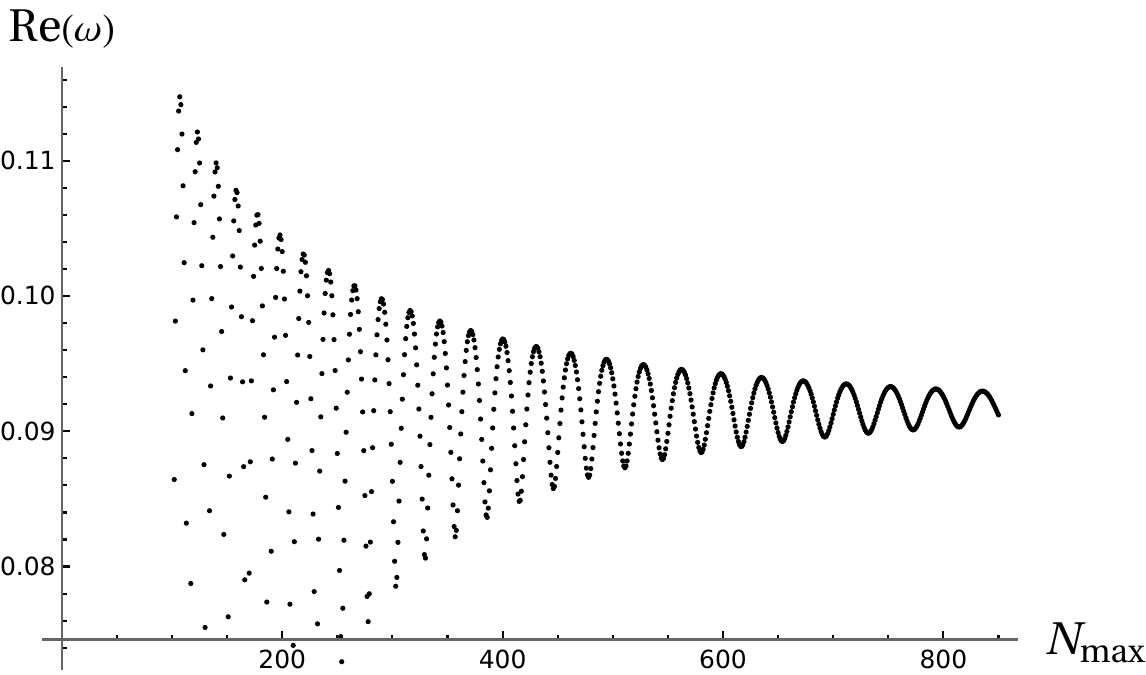}
\caption{The $(2+1)$-dimensional hydrodynamic black hole.  The real part of the
quasinormal frequency of the $m=1$, $n =2$ perturbation as a function of the
dimension of the Hill matrix..}
\label{fig6}
\end{figure}

\begin{figure}
\centering
\includegraphics[width=12.5 cm]{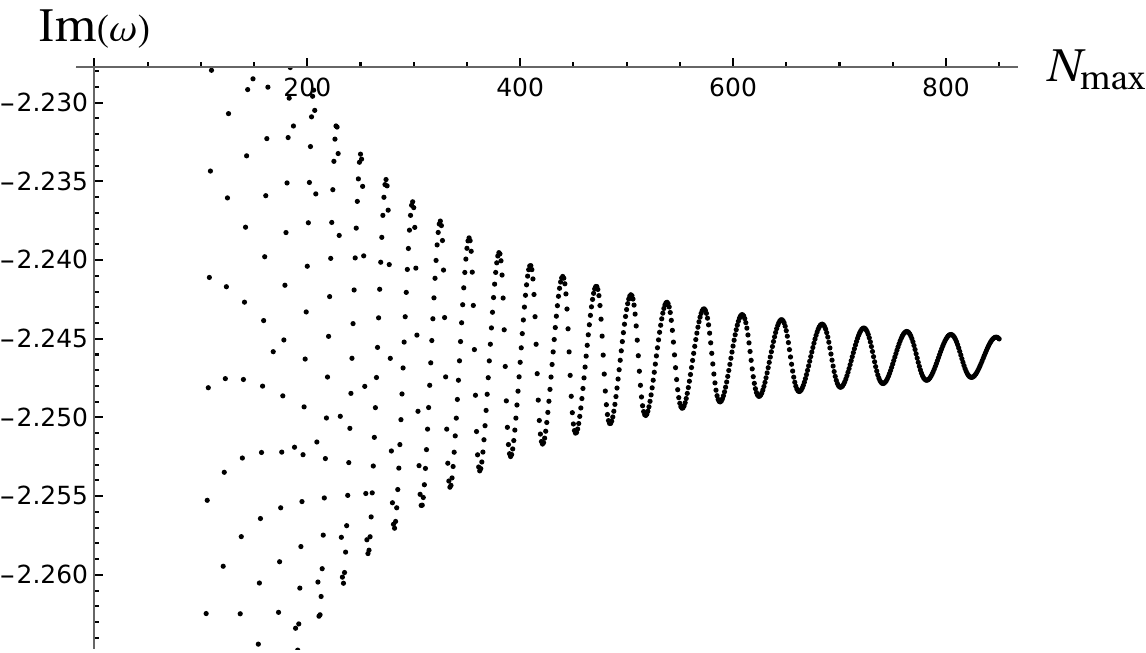}
\caption{The $(2+1)$-dimensional hydrodynamic black hole.  The imaginary part of the
quasinormal frequency of the $m=1$, $n =2$ perturbation as a function of the
dimension of the Hill matrix..}
\label{fig7}
\end{figure}

Results of our calculations are tabulated in Tab.~\ref{tab1}. The first two columns
specify the mode, $N_{max}$ is the dimension of the maximal matrix, $\omega$ is the
quasinormal mode rounded to 9 decimal places, and finally $\Delta \omega =
\omega_{Hill} - \omega_{WKB},$ where $\omega_{Hill}$ and $\omega_{WKB}$ are the
complex frequencies calculated within the framework of the Hill-determinant method
and the WKB-Pad\'e (or the WKB-Wynn method), respectively.  For each mode, we solved
the determinant equations
\begin{equation}
\det \mathcal{H}_{k} = 0
\end{equation}
for $6 \leq k \leq N_{max}$. Inspection of the table shows that the agreement between
the results obtained within the framework of the Hill and the WKB method is, in most
cases, truly remarkable. Except for the $(1,1)$, $(1,2)$, $(2,3)$, and $(3,4)$ modes,
where the first element of the ordered pair is $m$ and the second is $n,$  the real
and imaginary parts of $\Delta\omega$ are always smaller than $10^{-10}$. To
understand the origin of this (extremely small) discrepancy, let us analyze the
behavior of the roots of the determinant equations for the $(1,0)$, $(1,1)$, and
$(1,2)$ modes. Let us start with the fundamental one.  In this case (Fig.~\ref{fig0}),
convergence is rapid and the approximants are located along the spiral curve in the
complex $\omega$-plane.  It is noteworthy that satisfactory results can be achieved
even without convergence acceleration.  On the other hand, the
$(1,1)$-mode necessitates such an acceleration. Indeed,  as can be seen in
Fig.~\ref{fig1}, the approximants lie along the spiral curve, but now for $6 \leq k
\leq 850$ the convergence of $\omega_{k}$ is slower. (To enhance clarity, we included
additional 250 points).  Instead of further increasing $N_{max},$ we used the Wynn
acceleration technique and, after some computations, we arrived at the result quoted
in Tab.~\ref{tab1} and shown as a red dot in Fig.~\ref{fig1}.  The oscillatory character of
the real and imaginary parts of $\omega_{k}$ is clearly visible in Figs.~\ref{fig2}
and \ref{fig3}.  Despite the fact that $\Delta \omega$ does not satisfy our accuracy
criteria, we believe that the figures quoted in the table are correct, as they are
also confirmed by the analogous calculations done using the continued fraction
method.  Finally, let us analyze the $(1,2)$-mode. The results of our calculations
are shown in Fig.~\ref{fig4}. We see that the convergence is really slow and to gain a
better understanding of how successive  $\omega_{k}$ are arranged on a complex plane,
we connected the points corresponding to $k-1$ and $k$ by line segments
(Fig.~\ref{fig5}).  This explains why it is difficult to stabilize  the result to the
required precision, even after the applying the convergence acceleration algorithm.
Now, inspection of Figs.~\ref{fig6} and \ref{fig7} suggests that the exact value of
$\omega$ should be close to the value displayed in the table. Once again the quoted
$\omega$ has been compared with the analogous result calculated using the continued
fraction method.

%
%
\begin{table}
	\caption{Fundamental modes  of the  $(2+1)$ and $(3+1)$-dimensional
hydrodynamic black hole calculated within the framework of the sixth-order WKB
method without any convergence acceleration. Here, $l$ is either the multipole number $s$ or the azimuthal number $m$ }
\begin{ruledtabular}
\begin{tabular}{ccccc}
$l$ & $l = m, \,\, \omega_{(2+1)}$ & $l=s,\,\,\omega_{(3+1)}$ \\
\colrule
 $ 0 $ & $                   $ & $ 0.05820-0.87431 i $\\
 $ 1 $ & $ 0.42722-0.33011 i $ & $ 1.09712-0.39365 i $\\
 $ 2 $ & $ 0.95143-0.35304 i $ & $ 1.41382-0.70152 i $\\
 $ 3 $ & $ 1.46852-0.35248 i $ & $ 2.12332-0.61734 i $\\
 $ 4 $ & $ 1.97645-0.35296 i $ & $ 2.75319-0.61870 i $\\
 $ 5 $ & $ 2.48119-0.35319 i $ & $ 3.38007-0.61989 i $\\
 $ 6 $ & $ 2.98434-0.35331 i $ & $ 4.00536-0.62027 i $\\
\end{tabular}
\label{tab2}
\end{ruledtabular}
\end{table}

Let us compare our results with the results that can be found in the literature. In
Ref.~\cite{becarle}, the authors calculated quasinormal frequencies of the
fundamental modes $(n=0)$ for $1 \leq m \leq 4$ of the $(2+1)$-dimensional
hydrodynamic black hole using three simple WKB computational schemes.  The schemes differ only in
the number of terms retained in the right-hand-side of  Eq.~(\ref{main1}). The first
one was developed by Schutz and Will in which one retains only two first terms, and
their extensions based on the third-order $(N =3)$ and the sixth-order WKB approximation $(N=6)$. The
thus obtained results were rounded to two decimal places. To make the comparison more
accurate we have recalculated the fundamental modes for $1 \leq m \leq 6,$ using the
sixth-order WKB method independently and retained more decimal places.  The results
of our calculations are presented in Tab.~\ref{tab2}.  First, it should be noted that
our result for the mode (2,0) differs substantially from the analogous frequency
calculated in Ref.~\cite{becarle} and since our result has been confirmed by 
the two algorithmically different calculations
we believe that our result is correct. Now, we
can compare the results presented in Tab.~\ref{tab2} with the more accurate results
displayed in Tab.~\ref{tab1}.  As expected, larger discrepancies between our results
and those presented in Ref.~\cite{becarle} are observed in the lower modes and
the agreement improves progressively as the azimuthal number,  $m$, increases. This behavior
can be attributed to the increase of the real part of the complex frequencies of the fundamental modes
along with the relatively weak change in their imaginary parts.
 Typically, WKB-based methods 
for computing quasinormal modes (without application of any 
additional techniques, as, for example, convergence acceleration) 
are accurate for the frequencies satisfying $|\Re(\omega)| > |\Im(\omega)|$ 
and become less reliable when $|\Im(\omega)|$ increases. 

Specifically, for mode $(1,0)$ one has relative error 5\% and 3.3\%  for the real and
the imaginary parts, respectively, and the accuracy increases with $m.$
We have not attempted to make a similar comparison for
the overtones, but on general grounds,  we expect that the accuracy of the sixth-order WKB
rapidly deteriorates with the increase of $n.$ Indeed, for a given $m$ and increasing 
overtone number   $\Re(\omega)$ decreases and $|\Im(\omega)|$
increases, leading to a deterioration in the quality of the approximation.

Since the master equation of the $(2+1)$-dimensional analogue black hole is identical
to the equation describing the electromagnetic vector perturbation of the
five-dimensional Schwarzschild-Tangherlini black hole for $l\geq 2,$ it is possible
to extract valuable information from the calculations that have been carried out
earlier.  Indeed, it can easily be shown that the results presented here are consistent
with those  reported in Ref.~\cite{jose_5_dim1}. 
Moreover, a comparison of the complex frequencies listed in Tabs. VII - IX  of Ref.~\cite{ja_dim5} and Tab.~\ref{tab1} 
of the present paper shows perfect agreement. It should be emphasized, however, that the results 
presented here are a significant extension of the previous ones.

Finally, we make a few observations regarding the general behavior of the modes. Let us
start with the fundamental frequencies $(n=0)$. Inspection of Tab.~\ref{tab1} shows that their
real part increases whereas the imaginary part very slowly decreases wth $m$. For the first overtones
$(n = 1)$ the situation is different: both the real part and the imaginary part increases and
this trend also persists for higher overtones. On the other
hand, for a given azimuthal number $m$ the real and imaginary part of the complex
frequencies decreases.

\subsection{$(3+1)$-dimensional analogue black hole}

Let us consider the $(3+1)$-dimensional hydrodynamic black hole.  Unlike the previous
case of the $(2+1)$-dimensional black hole, there are no simple relations between $l$
and $s$ in Eqs.~(\ref{master1}) and (\ref{master}) with (\ref{hy4}),
respectively.  Since the application of the continued fraction method require, depending on the 
choice of the recurrence relation, either
five or three consecutive Gauss eliminations, we will not use this technique in this
case. Instead, we will apply the Hill determinant method for both recurrences
(\ref{rec2}) and (\ref{rec3}), and compare the thus obtained results with the
calculations carried out within the framework of the higher-order WKB method with the
convergence acceleration. Once again, we accept the thus obtained result  only if at
least two independent calculations agree to the required precision and the
calculations based on two different recursive schemes are considered as independent.

The results of the calculations are presented in Tab.~\ref{tab3}. Comparing the
performance of the methods we observe that the WKB is slightly less accurate,
particularly for the lowest fundamental mode. However, for the remaining modes the
agreement is really truly remarkable.  On the other hand, since for the $s = 0$ and
$n = 0$ mode both recurrences give the same result to the assumed accuracy, we
believe that the quoted figures are correct.  Now, we can compare our results with
the results discussed in Ref.~\cite{becarle}. Since the authors employed simple first, third,
and sixth-order WKB methods, the agreement of the first two fundamental modes is
poor. In particular, we do not confirm statement that for the fundamental $s = 0$ and
$s =1 $ modes the lowest WKB approximation gives the most reliable results.  We have
recalculated the sixth-order WKB results using our codes, extending the calculations
to $s = 6$  and retained more digits.  Inspection of the Tabs.~\ref{tab2}
and~\ref{tab3} shows, that for $s \geq 1$ it is the sixth-order WKB that agrees the
best with our results. On the other hand however, for the $(s = 0, n = 0)$ mode, all the
three simple approximations yield very poor results.  Since both the real and the modulus of
the imaginary part of $\omega$ rapidly grow in magnitude as the order of the WKB
increases, our conclusion is that it is not possible to obtain sensible results for
the $(s =0, n=0)$ mode without some modification of the method. 

Inspection of Tab.~\ref{tab3} shows that for the fundamental quasinormal modes both 
the real part of its frequency increases and the imaginary part slowly 
decreases as the $s$ number increases (an exception here is the $(0,0)$ mode). 
It seems that this weak dependence of $\Im(\omega)$ on $s$ is typical for both acoustic
black holes.
For the lowest overtones $(n=1)$ both real and imaginary part of $\omega$ increase.
On the other hand, for a given $s$ both the real part and the imaginary part of 
the frequency decrease.

\begin{figure}
\centering
\includegraphics[width=12.5cm]{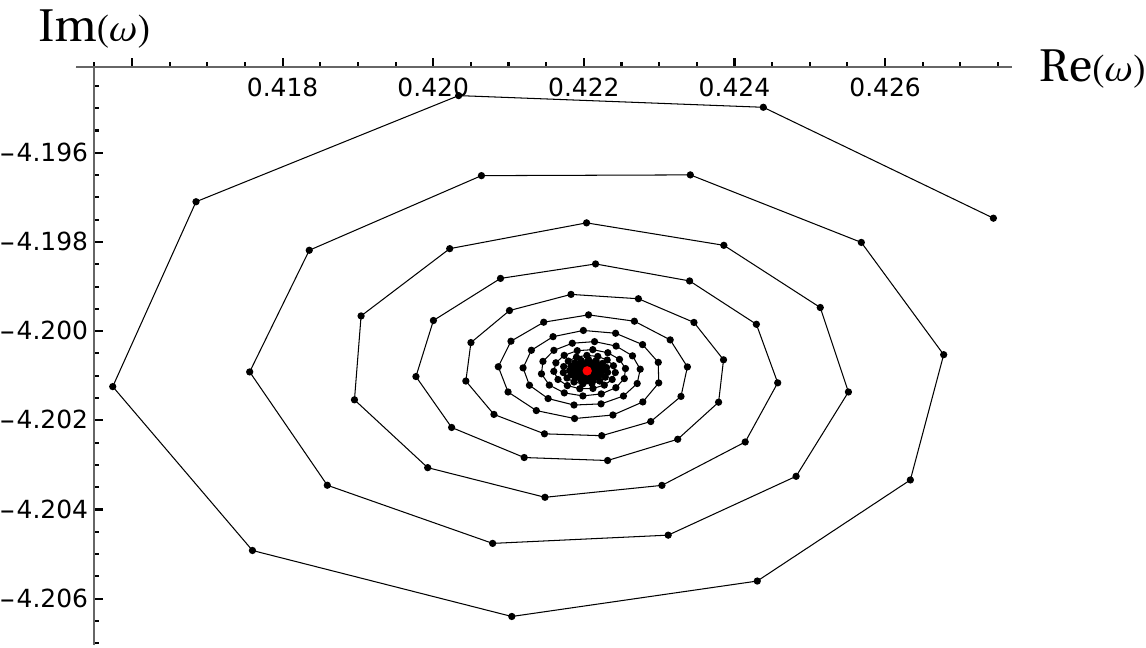}
\caption{The $(3+1)$-dimensional hydrodynamic black hole. Migration of the
approximants of the complex frequency of the $s=2$, $n = 2$ mode on the complex
plane. The red dot represents the limiting value calculated using the Wynn
acceleration algorithm. The $N$-th approximants in connected by a line segment with
the $N+1$.}
\label{fig8}
\end{figure}

\begin{figure}
\centering
\includegraphics[width=12.5 cm]{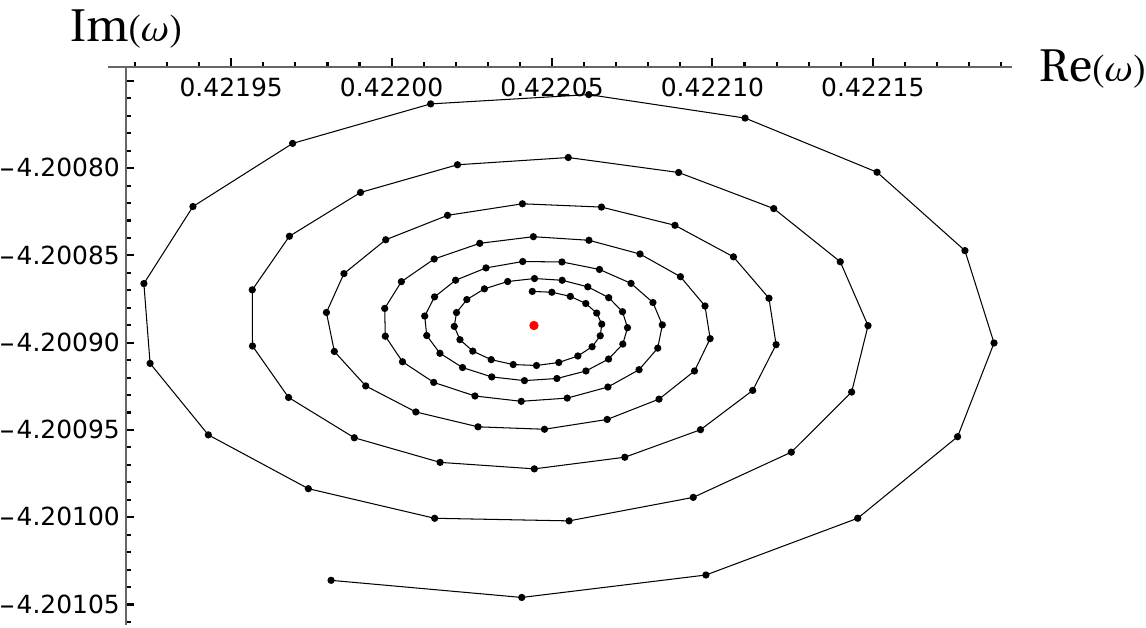}
\caption{The  central region of Fig.~\ref{fig8}.  
The red dot represents the limiting value calculated using the Wynn
acceleration algorithm.
}
\label{fig9}
\end{figure}

A typical migration of the consecutive approximations to the frequency of the 
quasinormal mode $s =2$ and $n=2$ is shown in Figs.~\ref{fig8} and \ref{fig9}.
Each point on the complex plane is a particular solution of the Hill determinant equations
calculated with the same initial condition. The red point represents 
the limit $\omega = 0.422044365 - 4.200890173 i.$
For a better visibility we have connected $k$ and $k-1$ points by  line segments,
where $k$ is the dimension of the Hill matrix. Since the solutions of the determinant
equation lie on the spiral-like curve the real and the imaginary parts of $\omega$ reveal 
the oscillatory character. This behavior resembles the behavior of the quasinormal modes
of $(2+1)$-dimensional analogue black holes.
\begin{table}
	\caption{Quasinormal modes of the $(3+1)$-dimensional hydrodynamic black
hole. $s$ is the multipole number, $n$ is the overtone number, $N_{max}$ is the maximal
dimension of the Hill matrix, $\omega$ is the complex frequency calculated within the
frameworks of the Hill determinant method and $\Delta \omega = \omega_{Hill} -
\omega_{WKB}$.  The high precision results of $\omega$ are rounded to 9 decimal
places.}
\begin{ruledtabular}
\begin{tabular}{ccccc}
$s$ & $n$ & $N_{max}$ & $\omega$  &$ \Delta \omega = \omega_{Hill} - \omega_{WKB}$  \\
\colrule
$ 0 $ & $ 0$ & $300$ & $ 0.072813815-0.619901186 i $ & $ 4.09 \times 10^{-5} -2.53 \times 10^{-5} i$ \\
$ 1 $ & $ 0$ & $200$ & $ 0.816775035-0.611054595 i$ & $-1.16 \times 10^{-15}-2.51\times 10^{-15} i $  \\
$ 2 $ & $ 0$ & $250$ & $1.479691724-0.617355106 i$ & $2.65  \times 10^{-22}+5.06\times 10^{-22} i $  \\
$  $ & $ 1$ & $300$ & $0.919814109-2.070617645 i$ & $1.91  \times 10^{-9}+ 2.45\times 10^{-10} i $  \\
$ 3 $ & $ 0$ & $200$ & $2.120221930-0.619367816 i$ & $-7.12 \times 10^{-25}+5.18\times 10^{-25} i $  \\
$  $ & $ 1$ & $250$ & $1.712558987-1.947807598 i$ & $4.46  \times 10^{-18}+1.67\times 10^{-19} i $  \\
$  $ & $ 2$ & $300$ & $0.984422651-3.769874878 i$ & $-7.03 \times 10^{-9}+1.59\times 10^{-8}   i $  \\
$ 4 $ & $ 0$ & $150$ & $2.752103594-0.620056709 i$ & $1.04  \times 10^{-23}+9.03\times 10^{-24} i $  \\
$  $ & $ 1$ & $200$ & $2.437907970-1.909000239 i$ & $-2.50 \times 10^{-21}-4.30\times 10^{-21} i $  \\
$  $ & $ 2$ & $250$ & $1.775391582-3.433842488 i$ & $-8.05 \times 10^{-14}+1.86\times 10^{-14} i $  \\
$  $ & $ 3$ & $300$ & $1.109803434-5.521445321 i$ & $6.01  \times 10^{-8}-2.68\times 10^{-8}   i $  \\
$ 5 $ & $ 0$ & $100$ & $3.379842300-0.620312172 i$ & $1.75  \times 10^{-21}-1.74\times 10^{-21} i $  \\
$  $ & $ 1$ & $150$ & $3.124541370-1.891903509 i$ & $-3.15 \times 10^{-20}-5.09\times 10^{-20} i $  \\
$  $ & $ 2$ & $200$ & $2.585373209-3.290233538 i$ & $-8.95 \times 10^{-18}+2.95\times 10^{-18} i $  \\
$  $ & $ 3$ & $250$ & $1.806285519-5.064977404 i$ & $4.61  \times 10^{-12}+5.12\times 10^{-12} i $  \\
$  $ & $ 4$ & $300$ & $1.259981716-7.272525138 i$ & $5.36  \times 10^{-8}+6.43\times  10^{-8}  i $  \\
\end{tabular}
\label{tab3}
\end{ruledtabular}
\end{table}

\section{Final remarks}
\label{sekcja7}

In this paper we have calculated the complex frequencies of the quasinormal
modes, $\omega,$ of the two models of the analogue black holes: the
$(2+1)$-dimensional draining bathtub black hole and the $(3+1)$-dimensional
canonical acoustic black hole. The perturbations satisfy the Schr\'odinger-type
differential equations and the complexity of the potential terms  effectively
preclude construction of the exact solutions.  Our calculations were performed
within the frameworks of the three major techniques: the Hill-determinant
method with the convergence acceleration, the continued fraction method also
with the convergence acceleration, and the higher-order WKB-Pad\'e method (or its modifications).
Each of the methods have its strong and weak points, and one of our principal
tasks was comparison their the effectiveness and performance. The other two tasks
were: calculation of the highly accurate $\omega$, say to 9 decimal places, and
extending (and sometimes correcting) the results found in the literature.
Moreover, knowledge of the highly accurate and well-documented quasinormal
modes may play a crucial role in the development of new computational methods and techniques.  

The signal from such hydrodynamic systems will certainly not be pure.
On the contrary, the signal from the man-made acoustic black hole will be
contaminated by other frequencies related to the way the black hole would
be excited, characteristics of the setup, and many other factors. It is therefore 
crucial to have solid theoretical predictions at one's disposal.

Now, we provide a brief characterization of each method in the context 
of the obtained results.  Our calculations indicate that
all of them work very well. The are however some differences in their performance.
Undoubtedly, the strength of the WKB-Pad\'e method lies in its black-box nature.
Once the potential $V(r)$ for a given $l$ and the overtone number is known, 
the application of the simple algorithm yields reasonable (in the most cases) results.
The limitations are typical for any method based on the WKB approximation: the results
deteriorate for higher overtones. It remains an open question whether this can be 
circumvented by increasing the number of the $\Lambda_{k}$ terms left in Eq.~(\ref{main1}).
We know that increasing the order of the WKB approximation results in rapid 
increase of the absolute values of the real and imaginary parts of the quasinormal 
frequency. So the above question is, in fact, a question regarding the 
effectiveness of the Pad\'e summation or other techniques for convergence acceleration. 

On the other hand, the Hill-determinant method is slightly more complicated. 
It requires knowledge of the analytical structure of the radial equation and 
the construction of appropriate recurrence relations for the numerical 
coefficients of the series expansion of its solutions. As the approximate 
quasinormal modes are a subset of the set of all solutions of the determinant 
equation, it is necessary to define some criteria for selecting solutions that 
are of interest to us. Our strategy is as follows: First, we calculate interesting 
mode using some low-cost method, and subsequently use it as a a starting point
for constructing the solution of the determinant equations for increasing 
dimension of the Hill matrices. The thus determined solutions migrate 
on a complex plane and in order to calculate the final result we accelerate 
convergence using the Wynn algorithm. This procedure yields very accurate 
results.

The third approach that has been used in this paper (but only in the case 
of the $(2+1)$-dimensional acoustic black hole) is the continued fractions
method. It relies on the same recurrence relations as the Hill-determinant
method and utilizes the deep relationship between three-term recursion and the
continued fractions. Since the recurrences considered in this paper involve
more terms than three, it is necessary to apply successive Gauss eliminations.
Now, the evaluation of the continued fractions leads to the rational functions,
which can be solved numerically for the characteristic frequencies of 
the quasinormal modes. Once again, to accelerate convergence of the results
we used the Wynn algorithm. In this form, the continued fractions method 
is comparable in effectiveness to the Hill-determinant method. However, 
there are two important differences. In the Hill determinant method, 
there is no need for transforming the matrix to the tridiagonal form, 
which considerably simplifies calculations. On the other hand, 
the continued fraction method has a mechanism for evaluating 
the tail terms, a feature that improves quality of the approximation.

Our calculations demonstrate that all three methods work very-well
and indicate that we have achieved assumed accuracy. Specifically, in the 
$(2+1)$-dimensional case both the continued fraction method and the Hill
determinant method always give the same results to (at least) 9 decimal places.
The performance of the higher-order WKB-Pad\'e method is also very good, 
as can be seen in the tables.
In the $(3+1)$-case, instead of employing the continued fraction method we used
two recurrence relations: the eight-term and the six-term ones.  
Once again, we have excellent agreement in the results obtained in each of these approaches. 
It should be emphasized that the convergence acceleration is an indispensable tool,
especially for the overtones and the methods presented here can easily be adapted 
to other types of black holes.
\begin{acknowledgments}
JM was partially supported by Grant No. 2022/45/B/ST2/00013 of the National Science
Center, Poland.
\end{acknowledgments}

\appendix*
\section{Recurrences}

The recurrence relations (\ref{rec1}), (\ref{rec2}), and (\ref{rec3}) have been
calculated for the analogue black holes.  They are, respectively, the special cases
of the recurrence relations of the five-dimensional and the seven-dimensional
Schwarzschild-Tangherlini black holes. Here, we collect the coefficients of the
general recurrence equation
\begin{equation}
  a_{k+1}\, \alpha_{k} + a_{k} \,\beta_{k} + a_{k-1}\,\gamma_{k}
+ a_{k-2} \,\delta_{k} + a_{k-3}\, \kappa_{k} + a_{k-4}\, \lambda_{k}+
 a_{k-5} \,\mu_{k} + a_{k-6}\, \nu_{k} = 0.
 \label{eq_a1}
\end{equation}
In the formulae below, $L = l(l+ D-3)$, $D$ is the dimension of the
Schwarzschild-Tangherlini black hole,  $T  =  1-j^{2},$   where $j$ is the type of
perturbation given in (\ref{typy_zab}), and finally $a_{k} =0$ for $k \leq -1.$ 
The coefficients of the
four-term recurrence ($\kappa_{k} = \lambda_{k} = \mu_{k} = \nu_{k} =0 $) are given
by

 \begin{equation}
 \begin{array}{lcl}
 \alpha_{k} &=& -2 (k+1) (k-2 \rho +1),\\
 \beta_{k} &=& 5 k^2+k (5-16 \rho )+L+16 \rho ^2-8 \rho +\dfrac{9 T}{4}+\dfrac{3}{4},\\
 \gamma_{k} &=& -4 k^2+8 k \rho -\dfrac{9 T}{2}+4, \\
 \delta_{k} &=& k^2-k+\dfrac{9 T}{4}-2.
  \end{array}
  \label{eq_a2}
\end{equation}
In the case of the seven-dimensional Schwarzschild-Tangherlini black hole one has
either the six-term recurrence relation with ($\mu_{k} = \nu_{k} =0$)
\begin{equation}
\begin{array}{lcl}
\alpha_{k} &=& -4 (k+1) (k-\rho +1), \\
\beta_{k} &=&  14 k^2+k (14-24 \rho )+L+12 \rho ^2-12 \rho +\dfrac{25}{4} T +\dfrac{15}{4},  \\
\gamma_{k} &=& -20 k^2+28 k \rho -8 \rho ^2-25 T+20, \\
\delta_{k} &=& 15 k^2-k (16 \rho +15)+4 \rho ^2+8 \rho +\dfrac{75 }{2} T-30, \\
\kappa_{k} &=& -6 k^2+4 k (\rho +3)-4 \rho -25 T+18 \\
\lambda_{k} &=& k^2-3 k+\dfrac{25}{4} T -4.\\
\label{eq_a3}
\end{array}
\end{equation}
or the eight-term recurrence
\begin{equation}
\begin{array}{lcl}
\alpha_{k} &=& -8 (k+1) (k-\rho +1)   ,\\
\beta_{k} &=&36 k^2-64 k \rho +28 k+2 L+28 \rho ^2-28 \rho +\dfrac{25}{2}T+\dfrac{15}{2} ,\\[1.5 ex]
\gamma_{k} &=& -72 k^2+4 k (34 \rho +9)-2 L-66 \rho ^2-32 \rho -\dfrac{125}{2}T+\dfrac{57}{2} ,\\[1.5 ex]
\delta_{k} &=& 84 k^2-152 k (\rho +1)+L+69 \rho ^2+142 \rho +\dfrac{525}{4} T-\dfrac{113}{4} ,\\
\kappa_{k} &=& -2 \left(31 k^2-k (49 \rho +97)+18 \rho ^2+80 \rho +75 T+12\right),\\
\lambda_{k} &=& 29 k^2-3 k (12 \rho +43)+9 \rho ^2+84 \rho +100 T+52 ,\\
\mu_{k} &=& -8 k^2+6 k \rho +46 k-18 \rho -\dfrac{75}{2}T-30,\\
\nu_{k} &=& k^2-7 k+\dfrac{25}{4}T+6 .
 \end{array}
 \label{eq_a4}
 \end{equation}
It should be noted that in general we do not restrict $j$ 
to the values given in (\ref{typy_zab}).

\clearpage


\end{document}